\documentstyle[12pt]{article}
\newcommand{\beq}{\begin{equation}}
\newcommand{\eeq}{\end{equation}}
\newcommand{\beqn}{\begin{eqnarray}}
\newcommand{\eeqn}{\end{eqnarray}}
\def\vdir{v\kern-7.8pt\Big{/}}
\def\pdir{p\kern-7.8pt\Big{/}}

\begin{document}

\title{Consistency and lattice renormalization of the
effective theory for heavy quarks}
\vskip 2.5truecm
\author{
U.~Aglietti\\
SISSA-ISAS, Via Beirut 2, 34014 Trieste, Italy \\
INFN, Sezione di Trieste, Via Valerio 2, 34100 Trieste, Italy\\}
\date{}
\maketitle
\begin{abstract}
\noindent

The effective theory describing infinite mass particles with a given
velocity, has a great interest in heavy flavor physics.
It has the unpleasant characteristic that
the energy spectrum is unbounded from below; this fact
is the source of the problems in the formulation of the euclidean
theory. In this paper we present an analysis of the euclidean
effective theory, that is rather complete and has positive conclusions.
A proof of the consistency of the euclidean theory is presented and
a technique for the evaluation of the amplitudes
in perturbation theory is described.
We compute also the one-loop
renormalization constants of the lattice effective theory and
of the heavy-heavy current that is needed for the determination
of the Isgur-Wise function.
A variety of effects related to the explicit breaking of the Lorentz
symmetry of lattice regularization is demonstrated.
The most peculiar phenomenon is that the heavy quark velocity
receives a finite renormalization.
Finally, we compute the lattice-continuum renormalization constant
of the Isgur-Wise current. It is needed for the conversion
of the values of the matrix elements computed with the lattice effective
theory, to the values in the full theory.
\end{abstract}

\newpage
\section{Introduction}

\noindent

A test of the Standard Model can be realized in the quark mixing
sector by measuring the entries of the CKM matrix and checking the
unitary relations.
The extraction of the values of the matrix elements from the
experimental data is possible only if one is able
to compute the effects of the strong interaction on the weak
process.
A first principle technique for the evaluation of the matrix
elements of the weak hamiltonian between hadron states is lattice
QCD \cite{mart}. Since present lattice cut-offs are at most $2\div 3~ GeV$,
it is not possible to simulate directly the dynamics of heavy quarks.
To circumvent this problem, effective theories have been constructed
in which the heavy quark masses are sent to infinity \cite{lt,ef,cl}.

A method to get a precise determination of $\mid V_{cb}\mid$
is offered by the analysis of the decays \cite{neu}:
\beqn
B\rightarrow D^{(*)}+l+\nu_l
\label{eq:dec}\eeqn
The hadronic matrix elements for the processes (\ref{eq:dec})
can be safely computed with lattice QCD by sending the beauty
and the charm masses to infinity: $M_b, M_c\rightarrow\infty$.
In this limit the matrix elements can be expressed
in terms of a single function, the Isgur-Wise function
$\xi$ \cite{vs,iwise}:
\beqn
\langle D,v\mid V_{\mu}(0)\mid B,v'\rangle &=&\sqrt{M_DM_B}~
(v_{\mu}+v'_{\mu})~\xi(v\cdot v')
\nonumber\\
\langle D^*,v,\epsilon\mid V_{\mu}(0)\mid B,v'\rangle &=&-i\sqrt{M_DM_B}~
\epsilon_{\mu\nu\alpha\beta}~\epsilon^{\nu}v'^{\alpha}v^{\beta}
{}~\xi(v\cdot v')
\nonumber\\
\langle D^*,v,\epsilon\mid A_{\mu}(0)\mid B,v'\rangle &=&\sqrt{M_DM_B}
(\epsilon_{\mu}(1+v\cdot v')-v_{\mu}v'\cdot\epsilon)
\xi(v\cdot v')~~~~~~
\label{eq:iwg}\eeqn
where $v'$ and $v$ denote respectively the $b$ and $c$ quark 4-velocities.

\noindent
$\xi(v\cdot v')$ is normalized at zero recoil, $\xi(1)=1$.

The computation of the Isgur-Wise function  with lattice QCD
requires the formulation of the theory of infinite mass quarks with
non zero velocity \cite{geo} on an euclidean lattice.

There are also other interesting applications.
The effective theory allows the
computation with lattice QCD of the production rate of heavy mesons
in $e^+e^-$ annihilations \cite{pair}:
\beqn
e^+e^-\rightarrow D^{(*)}+\overline{D}^{(*)},~~~B^{(*)}+\overline{B}^{(*)}
\eeqn
For center of mass energies far away from the masses of
the $c\overline{c}$ or $b\overline{b}$ resonances,
the heavy quarks are produced by the electromagnetic current
with velocities that are not appreciably changed by the hadronization.
The approximation of neglecting recoil effects for the heavy quark
dynamics is equivalent to the infinite mass limit.

Studies of the lattice
effective theories have already appeared in the literature. The
analytic continuation from minkowski to euclidean space has been
treated in ref.\cite{mo}. Specific properties of the
euclidean effective theories have been discussed in ref.\cite{noi}.
These peculiarities originate from
the fact that the energy spectrum of the effective theory is unbounded
from below; it is the expansion of the energy-momentum relation of a
heavy quark with momentum $\vec{p}~=~M\vec{v}+\vec{k}$ for small
$\vec{k}$:
\beqn
E~=~\sqrt{ M^2+\vec{p}^{~2} }~=~Mv_0+\vec{u}\cdot\vec{k}+
\frac{\vec{k}^2-(\vec{u}\cdot\vec{k})^2}{2M~v_0} +\ldots
\eeqn
where $\vec{u}$ is the kinematical velocity, $\vec{u}=d\vec{x}/dt=
                                              \vec{v}/v_0$.

\noindent
By removing the energy $Mv_0$ (it is a constant in a given velocity
sector) and neglecting $1/M$ terms, one gets the energy-momentum
relation of the effective theory,
\beqn
\epsilon~=~\vec{u}\cdot\vec{k},
\label{eq:neg}\eeqn
that is unbounded from below.

\noindent
The presence of states with negative energy is then an intrinsic
property of the effective theory and is related to the fact that one
removes the energy $Mv_0$ associated to a non zero 3-momentum
$M\vec{v}$. If the heavy quark picks up a residual momentum $\vec{k}$
with a component antiparallel to $M\vec{v}$, the energy decreases with
respect to $Mv_0$, and one is left with negative energies
in eq.(\ref{eq:neg}).
Nevertheless, the theory with $\vec{v}\neq 0$ is stable because it
is generated with a Lorentz transformation of the static theory,
and the latter has not negative energies ($\vec{u}=0$ in
eq.(\ref{eq:neg})).
The states with negative energies are simply an effect of the change
of reference frame and do not give rise to any instability.

The main consequences of the energy-momentum relation (\ref{eq:neg}),
as shown in ref.\cite{noi}, are:

\noindent
i) The free propagator of the heavy quark in space-time can be defined
only introducing an ultraviolet cut-off $\Lambda$ on the residual momenta
$\vec{k}$.

\noindent
ii) The correlators of systems composed of the effective quark and
light degrees of freedom reproduce correctly the lowest order in $1/M$
of the original theory.
Considering proper observables, it is also possible to take the continuum limit
$\Lambda\rightarrow\infty$.
These results are suggested by the explicit computation of a
heavy-light correlator in the free case.

\noindent
iii) Simple Feynman rules in 4-momentum space cannot be derived.

This paper  continues the analysis started in ref.\cite{noi} and  is
organized as follows.

\noindent
In sec.2 we report a computation  that includes the interaction among the
effective and the light quarks,
of the correlator considered in ref.\cite{noi} for the free case.
The results corroborate the conclusion of ref.\cite{noi}, namely that the
effective theory has a sensible continuum limit and correctly
reproduces the correlations of the full theory at lowest order in
$1/M$.

\noindent
In sec.3 it is presented a technique for computing amplitudes
of the euclidean effective theory in
perturbation theory.
Only slight modifications with respect to ordinary field theories
are needed.
One still has propagators and vertices and
the only difference with respect to an ordinary theory is that
it is not possible to
integrate any more the loops over real domains in momentum
space.
An $i\epsilon$ prescription is impossible,
as is instead the case for the static theory \cite{eh1},
 and one needs an additional
rule for constructing contours for the energy integration.

\noindent
Sec.4 deals with the lattice regularization of the effective
theory. The problem of the doubling of the heavy quark species is
discussed and the diagrams that are needed for the renormalization are
calculated.

\noindent
In sec.5 we discuss the renormalization of the lattice effective
theory. A computation of the renormalization constants at full order
$\alpha_S$ is presented.
We compute also the renormalization constant of a current
of the form $J=\overline{h}_{v}\Gamma h_{v'}$, where $h_{v},~h_{v'}$
are two heavy quark fields with 4-velocity $v$ and $v'$ respectively,
and $\Gamma$ is a generic Dirac matrix. The renormalization of this operator
is essential for converting the values of the Isgur-Wise function
computed with lattice QCD to the values in the full theory.

In general, the full order $\alpha_S$ renormalization on the
lattice shows a variety of effects related to the non-covariance of the
regularization. Mass and wave function
renormalization constants depend on the velocity of the heavy quark,
contrary to the case of a covariant regularization (like dimensional
regularization).
The renormalization constant of the current $J$ does not depend only
on $v\cdot v'$ (the only non trivial invariant given $v$ and $v'$),
but on the components of $v$ and $v'$ separately.

A very peculiar effect is also demonstrated, related to the fact
that the effective theory
contains an additional parameter with respect to the original theory,
namely the heavy quark 4-velocity $v^{\mu}$. There is a finite
renormalization of the velocity $\delta v$, that we have computed.
It is absent in a covariant regularization.
This effect does not spoil the effective theory of physical
meaning and the original normalization of the velocity is manteined,
i.e. $v_{R}^2~=~v_{B}^2~=~1$, where $v_{B}$ is the 'bare'
velocity and $v_R$ is the renormalized one $v_R=v_B+\delta v$.

In sec.6 we discuss the matching of the lattice effective theory with
the original high-energy theory.

Sec.7 contains the conclusions of our analysis.
\vskip .5truecm
\section{Consistency of the theory}

In this section we present a study of the consistency of
the euclidean effective theory.

We consider for simplicity a simple regularization with a cut-off
$\Lambda$ on the spatial momenta.
The free space-time propagator of a heavy quark in motion along the $z$
axis is given by \cite{noi} (see section 4 for a derivation):
\beqn
iH_{\Lambda}(t,\vec{x})&=&
\int^{\Lambda}
\frac{d^3k}{(2\pi)^3}
e^{ +i\vec{k}\cdot\vec{x} } iH(t,\vec{k})
\nonumber\\
&=&\frac{\theta(t)}{v_0}\delta(x)_{\Lambda}\delta(y)_{\Lambda}
\frac{1}{2\pi} \frac{e^{\Lambda(iz-u t)}-e^{-\Lambda(iz- u t)}}{ i z - u t }
\label{eq:cut}
\eeqn
where in the last line the velocity $\vec{u}$ has been taken along the
$z$ axis. $\delta_{\Lambda}$ is a regularized delta function.

\noindent
We consider a correlator
$G(t,\vec{k})$ of a system composed of an effective and a light
particle. Since what matters in this context
is the singularity structure of the amplitudes, let us consider
for simplicity scalar particles.
In the free case the correlator $G^{(0)}(t,\vec{l})$ is given by \cite{noi}
(see fig.1):
\beqn
G^{(0)}(t,\vec{l})&=&\int d^3 x e^{-i\vec{q}\cdot\vec{x}}
              iH(t,\vec{x}\mid 0)_{\Lambda}i\Delta_F(0\mid \vec{x},t)
\nonumber\\
&=& \Theta(t)e^{-\vec{u}\cdot\vec{l}~t}
\int^{\Lambda} \frac{d^3 k}{(2\pi)^3}\frac{1}{v_0 2 E(m\vec{v}-\vec{k}) }
e^{ -[\vec{u}\cdot\vec{k}+E(m\vec{v}-\vec{k})]~t  }~~~~~
\label{eq:g0}\eeqn
where $\Delta_F$ is the propagator of the light particle of mass $m$,
$\vec{P}$ is the total momentum of the composite system,
$\vec{q}=\vec{P}-M_Q\vec{v}$ is the momentum of the meson in the
effective theory and $\vec{l}=\vec{q}-m\vec{v}$
is the residual momentum.

\noindent
At large $\mid\vec{k}\mid$,
the argument of the exponential becomes
\beqn
\vec{u}\cdot\vec{k}~+~\mid \vec{k}\mid
\eeqn
and it is positive for $u<1$. The negative energies of the
effective particle are compensated by the positive energies of
the light particle in the states with high virtuality.
It is possible to take the continuum limit $\Lambda\rightarrow\infty$.
The argument of the exponential and the prefactor in eq.(\ref{eq:g0})
are the expansion of the full theory expressions,
and the Green function $G^{(0)}(t,\vec{l})$ correctly
describes the internal dynamics
of the composite system in lowest order in $1/M$.
At large times $t$ the correlator is dominated by the states
with the lowest invariant mass (that eventually become the
lowest bound state in the interacting theory), and it behaves like \cite{noi}:
\beqn
G^{(0)}(t,\vec{l}) \sim e^{-(mv_0+\vec{u}\cdot\vec{l})~t}
\label{eq:free}\eeqn
The $2$-point function $G^{(0)}(t,\vec{l})$
describes then a system with an infinite mass,
velocity $\vec{u}$, and residual momentum $\vec{l}$ (the energy
$Mv_0$ is removed), as the result of the correct coupling of a light
particle with an infinite mass particle.

We extend these results to the case of the interacting theory,
considering the exchange of a single scalar massless particle.
The correlator $G(t,\vec{l})$ is given at this order by:
\beqn
G^{(1)}(t,\vec{l}~)~=~G_a(t,\vec{l}~)+G_b(t,\vec{l}~)+G_c(t,\vec{l}~)
\eeqn
where the indices $a-c$ refer to the fig.2 at the end.

\noindent
By explicit computation one derives:
\beqn
G_a(t,\vec{l})&=&\int^{\Lambda} {\rm d}^3k{\rm d}^3p~\{
{}~A(\vec{k},\vec{p})\exp(~-[\vec{u}\cdot\vec{k}+E(m\vec{v}+\vec{l}-\vec{k})]t~)+
\nonumber\\
&+&B(\vec{k},\vec{p})
\exp(~-[\vec{u}\cdot(\vec{k}+\vec{p})+E(m\vec{v}+\vec{l}-\vec{k}-\vec{p})]t~)+
\nonumber\\
&+&C(\vec{k},\vec{p})
\exp(~-[\vec{u}\cdot(\vec{k}+\vec{p})+E(m\vec{v}+\vec{l}-\vec{k})+\mid
\vec{p}\mid]t~)+
\nonumber\\
&+&D(\vec{k},\vec{p})
\exp
(~-[\vec{u}\cdot\vec{k}+\mid\vec{p}\mid+E(m\vec{v}+\vec{l}-\vec{k}-\vec{p})]t~)
{}~\}
\label{eq:ga}\eeqn
where $\vec{p}$ is the momentum of the scalar particle
exchanged and $\vec{k}$
is the momentum of the heavy particle after the interaction.
The functions $A-D$ are very complicated functions of the loop
momenta $\vec{k}$ and $\vec{p}$ and we do
not report their expression. They are the correct expansion
of the corresponding functions of the full theory.

\noindent
For large loop momenta $\mid\vec{k}\mid$ and $\mid\vec{p}\mid$,
the arguments of the exponentials in the square brackets of eq.({\ref{eq:ga})
are positive and behave like:
\beqn
&& ~~~~~~~~~~~~~~~~\vec{u}\cdot\vec{k}+\mid\vec{k}\mid,~~~~~~~
   \vec{u}\cdot(\vec{k}+\vec{p})+\mid\vec{k}+\vec{p}\mid,
\nonumber\\
&& \vec{u}\cdot(\vec{k}+\vec{p})+\mid\vec{k}\mid+\mid\vec{p}\mid,
{}~~~~~~
\vec{u}\cdot\vec{k}+\mid\vec{p}\mid+\mid\vec{k}-\vec{p}\mid
\eeqn
The mechanism of compensation
of the energies works also in the interacting theory: the negative
energies of the heavy 'quark' are compensated by the positive
energies of the light 'quark' and/or the 'gluon'.
It is then possible to take the
continuum limit without encountering other divergences than those of
a usual field theory.
The functions $A$ and $B$ are multiplied by the exponentials with $2$
energies, that depend respectively only on $\vec{k}$ and
$\vec{k}+\vec{p}$. The divergences for $\Lambda\rightarrow\infty$
originate from the integration of $A$ over $\vec{p}$, and of $B$ over
$\vec{k}-\vec{p}$.

At large times $t$ the correlator (\ref{eq:ga})
behaves like the free one (\ref{eq:free}),
apart from ultraviolet divergences in the vertex
that can be factorized in the usual renormalization constants.

For the amplitudes (b) and (c) the same considerations as for the
amplitude (a) hold.
The explicit expression of $G_b$ is:
\beqn
G_b(t,\vec{l})&=&
\int {\rm d}^3k{\rm d}^3p        \{~
E(\vec{k},\vec{p})
\exp
(~-[\vec{u}\cdot\vec{k}+\mid\vec{p}-\vec{k}\mid+E(m\vec{v}+\vec{l}-\vec{p})]t~)+
\nonumber\\
&+&[F(\vec{k},\vec{p})+~tG(\vec{k},\vec{p})]
\exp(~-[\vec{u}\cdot\vec{p}+E(m\vec{v}+\vec{l}-\vec{p})]t~)   ~\}
\label{eq:gb}\eeqn
where $\vec{p}$ is the momentum of the heavy quark before the emission
of the scalar, and $\vec{k}$ the momentum of the heavy quark after
the emission.
The terms with $F$ and $G$ are associated respectively to the mass
and wave function renormalization of the heavy quark and will be
computed more easily in sec.4. $G_c$ is given by:
\beqn
&&G_c(t,\vec{l})=\int {\rm d}^3 k{\rm d}^3 p~\{~
M(\vec{k},\vec{p})\exp(~-[\vec{u}\cdot\vec{k}
+E(m\vec{v}+\vec{l}-\vec{k}-\vec{p})+\mid\vec{p}\mid ]t~)
\nonumber\\
&&+[N(\vec{k},\vec{p})+~tP(\vec{k},\vec{p})]
\exp(~-[\vec{u}\cdot\vec{k}+E(m\vec{v}+\vec{l}-\vec{k})
]t~)~\}
\label{eq:gc}\eeqn
where $\vec{k}$ is the momentum of the heavy quark and $\vec{p}$
is the momentum of the 'gluon'.

In lattice regularization the formulas (\ref{eq:ga}-\ref{eq:gc})
have obvious modifications;
the energies of the particles are replaced by the energies in lattice
regularization. One can easily check that the mechanism of compensation
of the energies is not spoiled by lattice effects.

Even though the computation that we have presented takes into account
the interaction only
at the lowest order, we argue that the results
shown have a general validity. Field fluctuations do not couple
to the negative energies, that
have a kinematical origin and do not point to any inconsistency.

\section{Contour representation of the amplitudes}

In this section we derive the rules for computing amplitudes
of the euclidean effective theory in perturbation theory.
A continuum regularization with a cut-off $\Lambda$ on the
spatial momenta is assumed for illustrative purposes;
the variations for the lattice case are straightforward and will
be discussed in sec.4.

As a first step, let us derive a contour representation of the
euclidean effective propagator.

\noindent
The correct propagator of the heavy quark $H(t,\vec{k})$,
as a function of time $t$
and spatial momentum $\vec{k}$, has been derived in ref.\cite{noi}:
\beqn
iH(t,\vec{k})~=~\frac{\Theta(t)}{v_0}e^{-~\vec{u}\cdot\vec{k}~t}
\label{eq:noi}\eeqn
It is forward in time, since it has to describe particle propagation
only, and contains the correct energy-momentum relation (\ref{eq:neg}).
Because of the exponential increase with time associated with negative
energy states $\vec{u}\cdot\vec{k}<0$,
the propagator (\ref{eq:noi}) cannot be represented as the Fourier
transform of a 4-momentum propagator.
By allowing the euclidean energy of the heavy quark $k_4$ to be complex,
one can write:
\beqn
iH(t,\vec{k})~=~\int_C \frac{dk_4}{2\pi}
             ~\frac{\exp(ik_4t)}{ iv_0k_4+\vec{v}\cdot\vec{k} }
\label{eq:contour}\eeqn
where the contour $C$ approaches the real line for
$k_4\rightarrow\pm\infty$, is oriented in the same way,
and passes {\it below} the singularity of
the integrand, at $k_4=i\vec{u}\cdot\vec{k}$, for {\it every sign} of
the energy.
For positive energies $C$ can be chosen as the real axis, and in this
case formula (\ref{eq:contour}) reduces
to the Fourier transform, while for negative
energies $C$ has to be moved in the lower half plane.
The representation (\ref{eq:contour})
can also be derived by means of a Wick rotation
in the complex plane of $k_0$, the energy in Minkowski space.
The propagator of the heavy quark in Minkowski space is given by:
\beqn
iH(k_0,\vec{k})~=~\frac{i}{v_0k_0-\vec{v}\cdot\vec{k}+i\epsilon}
\eeqn
There is a pole in the lower half plane,
at $k_0=\vec{u}\cdot\vec{k}-i\epsilon$. To preserve the causal
structure of the theory the Wick rotation has to be done
without crossing the pole; for positive energies the pole stays
in the right quadrant, and one can rotate the axes as in ordinary
field theories. For negative energies the pole stays in the left
quadrant, the rotation of the real axis
has to be accompanied by a deformation, and this produces the contour
in eq.(\ref{eq:contour}).

\noindent
In the static case $\vec{u}=0$, the pole of the integrand
in eq.(\ref{eq:contour})
stays in the real axis and one can impose the correct analytic
structure by an $i\epsilon$ prescription \cite{eh1}. The $i\epsilon$
of the static euclidean theory then has the meaning of a small
mass or positive kinetic energy to ensure the decay of the
correlations.

Let us consider now an amplitude containing a heavy quark propagator,
for example, the self-energy graph of fig.4
In Minkowski space the amplitude is proportional to
\beqn
\int \frac{d^4k}{(2\pi)^4} ~\frac{1}{v\cdot(p+k)+i\epsilon}
                      ~\frac{1}{k^2-\lambda^2+i\epsilon}
\label{eq:mink}\eeqn
where $p$ is the external momentum.

\noindent
In the lower half-plane of $k_0$ there are the heavy quark and the
gluon pole, at $k_0=-p_0+\vec{u}\cdot(\vec{p}+\vec{k})-i\epsilon,
                    ~E_{\lambda}-i\epsilon$,
while in the upper half plane there is the 'antigluon' pole,
at $k_0=-E_{\lambda}+i\epsilon$, where
$E_{\Lambda}=\sqrt{\vec{k}^2+{\lambda}^2}$ (see fig.3).
The real line separates the poles of the particles from the poles
of the antiparticles. In making the Wick rotation one has to
deform the real axis in order to keep the same topology.
The euclidean amplitude is then proportional to
\beqn
\int_C
\frac{d^4k}{(2\pi)^4}~\frac{1}{iv_0(k_4+p_4)+\vec{v}\cdot(\vec{p}+\vec{k})}
                     ~\frac{1}{k^2+\lambda^2}
\label{eq:eucl}\eeqn
where the contour $C$ divides the $k_4$-plane in two regions, one containing
the gluon pole and the heavy quark pole, the other containing the
'antigluon' pole.
The integration over $k_4$ of (\ref{eq:eucl}) gives:
\beqn
\frac{1}{v_0}\int \frac{d^3k}{(2\pi)^3}~
      \frac{1}{    [ip_4+E_{\lambda}+\vec{u}\cdot(\vec{k}+\vec{p})]
                         2E_{\lambda}   }
\eeqn
where the integration extends now to the ordinary 3-momentum space.

\noindent
By performing the integration over $k_0$ in (\ref{eq:mink})
one gets:
\beqn
\frac{i}{v_0}\int \frac{d^3k}{(2\pi)^3}~
      \frac{1}{    [-p_0+E_{\lambda}+\vec{u}\cdot(\vec{k}+\vec{p})]
                         2E_{\lambda}     }
\eeqn
The
two amplitudes are the correct continuation one of the other, i.e.
they give the same function of the external momentum $p^{\mu}$
if one sets $p_4= ip_0$.

{}From the above example it is easy to derive the general rule for
the contour of integration $C$ of the euclidean energy $k_4$:
$C$ must divide the $k_4$ plane into two connected regions,
one containing only the poles of the full and the effective
particles, the other containing the antiparticles poles.
To satisfy this requirement the contour $C$ has to be deformed
during the integration over the spatial momenta.

\noindent
For positive energies of the heavy quark the contour of
integration of $k_4$ can be chosen as the real axis, and for negative
energies the topology must remain the same.
The rule above can also be formulated in the following way:
one has to
integrate $k_4$ over the real axis by assuming that the poles
of the effective particles stay {\it always} in the region they
occupy for positive energies.

\section{Lattice regularization}

We assume the regularization of the euclidean effective theory that
has been proposed in ref.\cite{mo}, that is forward in time and symmetric
in space. Considering for simplicity a motion of the heavy quark
along the $z$ axis, the action $S$ is given by:
\beqn
iS &=&  -\sum_x
{}~v_0\psi^{\dagger}(x)~[\psi(x)-U_t^{\dagger}(x)\psi(x-\vec{t})]+~
\nonumber\\
 & &~~~~~-i\frac{v_z}{2}\psi^{\dagger}(x)~[U_z(x+\vec{z})\psi(x+\vec{z})
                                 -U_z^{\dagger}(x)\psi(x-\vec{z}) ]
\label{eq:act}\eeqn
where $\vec{\mu}$ is a versor in the direction $\mu$, and $U_{\mu}(x)$
are the links related to the gauge field by:
$U_{\mu}(x)~=~\exp[-igA_{\mu}(x-\vec{\mu}/2)]$.

Let us discuss the problem of the doubling of the heavy quark
species.

\noindent
The energy-momentum relation of the heavy quark on the lattice
is derived by computing the propagator as a function of time $t$ and
the residual momentum $\vec{k}$:
\beqn
iH(t,\vec{k})=\int_C \frac{dk_4}{2\pi}
                     \frac{e^{ik_4t}}{v_0(1-e^{-ik_4})+v_z\sin k_z}
          = \frac{\theta(t)}{v_0}e^{-(t+1)\ln(1+u_z\sin k_z)}~~~~
\eeqn
One has then:
\beqn
      \epsilon=1+u_z\sin k_z
\eeqn
The energy is zero not only at $k_z=0$ but also at $k_z=\pi$,
implying that the lattice regularization has produced a
duplication of the low energy excitations.
A regularization that is forward in time and in space is also
affected by the doubling problem. In this case the propagator is
given by:
\beqn
iH'(k)&=&\frac{1}{v_0(1-e^{-ik_4})~-iv_z(1-e^{-ik_z})}
\eeqn
and  the energy-momentum relation is:
\beqn
\epsilon ' ~=~\ln [ 1-iu_z(1-e^{-ik_z})]
\eeqn
The energy is a complex function of $\vec{k}$ and the doubling occurs
when $\epsilon '$ is purely imaginary, at:
\beqn
\cot(k_z/2)=-u_z
\eeqn

We show now by a physical argument that the doubling
has not any significant effect in the phenomenological applications
of the effective theory.
Consider a meson composed of a effective quark $Q$ and a light
antiquark $\overline{q}$, with
total momentum $\vec{P}$.
We assume that the doubling has been removed for the light quark.
The effective theory deals with the residual momentum
$\vec{k}$ of the meson:
\beqn
\vec{k}=\vec{P}-M_Q\vec{v}
\eeqn
where $M_Q$ is the heavy quark mass.

\noindent
It holds:
\beqn
\vec{k}=\vec{k}_Q+\vec{k}_l
\eeqn
where $k_Q$ and $k_l$ denote respectively the momentum of $Q$ and
of the light degrees of freedom.

\noindent
Since the large mass scale $M_Q$ is removed,
one expects, after renormalization,  $\mid\vec{k}\mid$ to be
of the order of the hadronic scale
$\Lambda_{QCD}$, that is much less than the lattice cut-off
$1/a$.

\noindent
This implies that when
$(k_{Q})_z=\pm\pi/a$ and the energy of the effective quark is zero,
the light quark momentum $\vec{k}_l$ is very near
to the ultraviolet cut-off and its kinetic energy
is very large, of the order of $1/a$.
The configurations in which the heavy quark has a momentum
at the edge of the Brillouin zone are then suppressed because of the large
energy of the heavy-light system, as it should be.
The situation is similar to that of a light meson composed of
a Wilson fermion ($r\neq 0$) and a naive fermion ($r=0$).
For small meson momenta $\mid\vec{P}\mid\ll 1/a$,
the internal dynamics is described correctly,
even though there is a duplication of the meson species.

The doubling has a negligible effect also in the dynamics of
the transition of a heavy meson into an heavy meson
(the dynamics of the Isgur-Wise form factor).
Due to the change of velocity of the
heavy quark after $W$ emission, the typical momentum
transfer $q^{\mu}$ between the heavy quark
and the light degrees of freedom may be greater than $\Lambda_{QCD}$.
By dimensional arguments one expects $q^{\mu}\sim\Lambda_{QCD}v\cdot v'$.
If also this scale is assumed to be much less than $1/a$, the
typical momentum exchanges lie in a region in which lattice effects
are negligible.

Assuming a convention for the Fourier transform according to which
$\psi(x)\sim\exp(ik\cdot x)$, one derives from the action
(\ref{eq:act}) the following Feynman rules:
\beqn
iH(k)&=&\frac{1}{v_0(1-e^{-ik_4})~+~v_z\sin k_z}
\\
V_0&=&i~g~v_0~t_a~e^{-i(k_4+k'_4)/2}
\\
V_z&=&g~v_z~t_a~\cos(k_z/2+k'_z/2)
\nonumber\\
V_0^{tad}&=&-\frac{g^2~v_0}{2}~t_at_b~e^{-ik_4}
\\
V_z^{tad}&=&\frac{g^2~v_1}{2}~t_at_b~\sin k_z
\nonumber\eeqn
where $k$ and $k'$ denote respectively the momenta of the incoming
and outcoming heavy quark, $V_0$ and $V_z$ are the interaction
vertices of the heavy quark with a gluon with a polarization along
the time or the $z$ axis. $V^{tad}$ are the vertices of emission of
two gluons, for the case of the tadpole graph ($k=k'$).

\noindent
It is to note that the conventions for the sign of the Fourier transform
and of the velocity are not independent, if one wants to interpret
$k$ as the residual momentum of the heavy quark.
If one assumes a convention according to which
$\psi(x)\sim\exp(-ik\cdot x)$, only the sign of $k_4$ changes in the
above Feynman rules (i.e. the sign in front of $v_z$ in eq.(\ref{eq:act})
is changed).

In usual lattice field theories every component of the loop momentum
$k_{\mu}$
is integrated in the interval $[-\pi,+\pi]$, i.e. $\exp(ik_{\mu})$
is integrated along a unitary circle.
For the effective theory the integration contour of $\exp(ik_4)$
has to be distorted in order to keep the poles of the
effective quarks always in the right region of the $\exp(ik_4)$-plane,
the region of the full particle poles.
The rule for the contour in lattice regularization
is analogous to the case of the continuum discussed in sec.3.

We apply now these rules to the calculation of the amplitudes that
are needed for the renormalization of the effective theory.
The infrared divergences are regulated by a fictious gluon mass
$\lambda$.

\noindent
The self-energy graph of fig.4 is given by:
\beqn
A(p)=-g^2C_F\int~\frac{d^4k}{(2\pi)^4}
\frac{v_0^2e^{-i(2p_4+k_4)}-v_z^2\cos^2(p_z+k_z/2)}
     {v_0(1-e^{-i(k_4+p_4)})+v_z\sin(k_z+p_z)}~
                              \frac{1}{\Delta(k)}
\eeqn
where the integration region is the domain $[-\pi,+\pi]^3\times C$.
$C_F=\sum t_at_a=(N^2-1)/2N$ for an $SU(N)$ gauge theory,
and $\Delta(k) = 2\sum (1-\cos k_{\mu}) +(a\lambda)^2$.

\noindent
Since this integral has to be computed numerically, it is convenient to
reduce the integration region to a real domain. Making
the contour integration analytically, one gets:
\beqn
A(p)~=~\frac{-g^2C_F}{16\pi^2}\frac{1}{\pi}
        &&\int d^3k\frac{1}{\sqrt{(1+A)^2-1} }\times
\nonumber\\
      &&\times \frac{v_0^2~z(k)~e^{-2ip_4}-v_z^2\cos^2(p_z+k_z/2)}
       {v_0(1-z(k)e^{-ip_4})+v_z\sin(k_z+p_z)}
\label{eq:selfc}\eeqn
where $A=\sum_{i=1}^3 (1-\cos k_i) +\lambda^2/2~$ and
$z(k)= 1+A-\sqrt{(1+A)^2-1}$

\noindent
The tadpole graph of fig.5 is given by:
\beqn
T(p)~=~\frac{-g^2C_F}{16\pi^2}~(v_0e^{-ip_4}-v_z\sin p_z)~
     \frac{1}{2\pi^2}\int \frac{d^4k}{(2\pi)^4}\frac{1}{\Delta (k)}
\eeqn
In this case there is no need to integrate over $k_4$ because the
integrand does not contain any effective propagator and the
integration region reduces to the ordinary one, $[-\pi,+\pi]^4$.

The vertex correction of the local heavy-heavy current
$J(x)=\overline{h}_{v}(x)\Gamma h_{v'}(x)$, omitting the trivial
spin structure $(1+\hat{v}')/2~\Gamma~(1+\hat{v})/2$, is given by (see fig.6):
\beqn
\delta V=-g^2C_F&&\int\frac{d^4k}{(2\pi)^4}
            \frac{1}{\Delta (k)}\times
\nonumber\\
&&\times
\frac{v_0v_0'e^{-ik_4}-v_zv_z'\cos^2(k_z/2)}{[v_0(1-e^{-ik_4})+v_z\sin k_z]
                                              [v_0'(1-e^{-ik_4})+v_z'\sin k_z]}
{}~~~~~\eeqn
where we have taken the motion of the two heavy quarks along the $z$
axis, and we have set to zero the external momenta.

\noindent
Integrating over $k_4$ one gets:
\beqn
\delta V=\frac{g^2C_F}{16\pi^2}\frac{-1}{\pi v_0v'_0}
&&\int\frac{d^3k}{ \sqrt{(1+A)^2-1} }\times
\nonumber\\
&&\times\frac{v_0v_0'z(k)~-~v_zv_z'\cos^2(k_z/2)}
            {(1-z(k)+u_z\sin k_z)(1-z(k)+u_z'\sin k_z)}
\label{eq:iwvert}\eeqn

\section{Lattice renormalization}

In this section we describe the one-loop renormalization of the
effective theory on the lattice.

The self-energy $\sum (k,v)$ of the heavy quark is given by the sum
of the graphs considered in section 4:
\beqn
\sum(k,v) ~=~ A(k,v)~+~T(k,v)
\eeqn
The bare propagator is given by:
\beqn
iH(k)~=~\frac{1}{v_0(1-e^{-ik_4})+v_z\sin k_z+M_0-\sum(k,v)}
\eeqn
where we have inserted a bare mass term $M_0$ to compute the mass
renormalization condition.

\noindent
We impose on-shell renormalization conditions. Near the mass-shell
the propagator looks like:
\beqn
iH(k)~=~\frac{1}{(iv_0-X)k_4+(v_z-Y)k_z+ M_0-\sum(0)+O(k^2)}
\label{eq:bare}\eeqn
where
\beqn
X=\left(\frac{\partial\sum}{\partial k_4}\right)(0),~~~~
Y=\left(\frac{\partial\sum}{\partial k_z}\right)(0)
\eeqn
Because of lattice effects (see later), the vector $(X,Y)$ turns out to be
not proportional to
the euclidean velocity $(iv_0,v_z)$.
This implies that mass and wave function renormalizations are not
sufficient for a complete renormalization of the effective theory.
This effect can be interpreted as a renormalization of the velocity.
The velocity $v$ appearing in eq.(\ref{eq:bare}) has to be identified with a
'bare' velocity $v_B$, that is modified by the field fluctuations
into a 'renormalized' velocity $v_R =v_B+\delta v$.
By comparing the bare propagator (\ref{eq:bare})
to the expression in terms of the renormalized parameters
\beqn
\frac{Z}{i(v_R)_0k_4 +(v_R)_zk_z + M_R+O(k^2)},
\eeqn
and imposing
the normalization of the velocity
\beqn
     (v_R)^2~=~(v_B)^2~=~1,
\eeqn
one gets, up to order first order in $\alpha_s$:
\beqn
\delta M &=& -\sum(0)
\label{eq:delm}\\
\delta Z&=& -iv_0X-v_zY
\label{eq:delz}\\
\delta v_z &=& -iv_0v_zX-v_0^2Y
\label{eq:delv}\eeqn
where $\delta Z= Z-1$.

The explicit expression for the mass renormalization $\delta M$
 is:
\beqn
\delta M ~=~
\frac{g^2C_F}{16\pi^2}[& &\frac{1}{\pi v_0}\int d^3k
\frac{v_0^2~z-v_z^2\cos^2(k_z/2)}{\sqrt{(1+A)^2-1}~(1-z+u_z\sin k_z)}+
\nonumber\\
& &+\frac{v_0}{4\pi^2}\int d^4k\frac{1}{\Delta(k)}~]
\eeqn
The first term comes from the amplitude (\ref{eq:selfc})
and has a relativistic invariant
form for small $k$; it is a function of the velocity because of hard
gluons. The second term originates
from the tadpole graph and is then a lattice effect.
It is also a function of the velocity
because of the explicit factor $v_0$.

The mass renormalization $\delta M$ is a function of the velocity
$u_z$. It is linearly divergent with the ultraviolet cut-off
$1/a$ and can be written as:
\beqn
\delta M ~=~ \frac{g^2C_F}{16\pi^2}~\frac{x(u)}{a}
\eeqn
The numerical values of $x(u)$ are reported in table at the end.
The numerical error is at most one unit in the second decimal place.
For $u=0$ one recovers the static value
already computed in ref.\cite{fr,eh2}.
At $\beta=6$ the mass renormalization is about 17\% of $1/a$ for
$u=0$ and decreases up to 9\% at $u=0.7$.

We note that the mass renormalization $\delta M$ in the effective theory
with $\vec{v}\neq 0$ is in effect a renormalization of the residual
momentum $k^{\mu}$ of the heavy quark. Indeed the heavy quark
propagator can be written in the limit $a\rightarrow 0$ as:
\beqn
\frac{1}{v\cdot k+\delta M}=\frac{1}{v\cdot (k-\delta M v)}
\eeqn
where $v$ is the euclidean $4$-velocity, $v=(iv_0,\vec{v})$
and $v_0=\sqrt{ 1+\vec{v}^{~2} }$.

\noindent
One can restore the original form of the propagator,
that has a pole at $k=0$, by
defining a renormalized residual momentum $k_R$ by means of the relation:
\beqn
k_R~=~k-\delta M v
\label{eq:krin}\eeqn
This effect has a very physical interpretation. The mass renormalization
of the static quark is given by the energy of the Coulomb-like field
surrounding the colour charge. For $\vec{v}\neq 0$, the Coulomb field
moves rigidly with the source, and carries the $3$-momentum
$\delta M\vec{v}$ in addition to the energy $\delta M v_0$.
If one wants to interpret $k$ as the fraction of the heavy
quark momentum that is changed in the collisions, and that is zero
in the absence of interactions with other particles, it is
necessary to subtract the constant contribution from mass
renormalization. In practise, it is not necessary to make the
subtraction (\ref{eq:krin}), because in the effective theory the
energy-momentum relation is linearized, and it does not matter
if the expansion point is shifted by renormalization.
The only effect of $\delta M v$ is an additional constant decay
of the propagator with time, according to
\beqn
iH(t,\vec{k})~\sim~e^{-(\delta M/v_0+\vec{u}\cdot\vec{k})t}
\eeqn
instead of
\beqn
iH(t,\vec{k})~\sim~e^{-\vec{u}\cdot\vec{k}~t}~~~~
\eeqn

According to eq.(\ref{eq:delz}) the expression
for the renormalization constant of the field $\delta Z$ is:
\beqn
&&\delta Z = \frac{g^2C_F}{16\pi^2}
{}~\frac{1}{\pi}~[~\int\frac{d^3k}{\sqrt{(1+A)^2-1}}
           \frac{2v_0^2~z(k)+v_z^2u_z\sin k_z}{1-z(k)+u_z\sin k_z}+
\nonumber\\
        &&~+\int \frac{d^3k}{\sqrt{(1+A)^2-1}}~
\frac{[v_0^2~z(k)-v_z^2\cos^2(k_z/2)][z(k)-u_z^2\cos k_z]}
     {[1-z(k)+u_z\sin k_z]^2}~]~~~~~
\label{eq:delz2}\eeqn
The first term in eq.(\ref{eq:delz2}) is infrared finite and comes from
the differentiation of the momentum-dependent vertices.
The second term is infrared divergent,
and the singularity
is isolated with the technique introduced in ref.\cite{bsd}; the remaining
integral is evaluated numerically. Details are given in appendix A.

\noindent
One can write:
\beqn
Z(u) ~=~ 1~+~\frac{g^2C_F}{16\pi^2}~[~-2\ln(a\lambda)^2 ~+~e(u)~]
\label{eq:delz3}\eeqn
The coefficient of the logarithmic term, i.e. the anomalous dimension
of the heavy quark field, is independent of the velocity.
It is indeed the same in every regularization and does not depend
on the velocity in a covariant regularization \cite{pw}.
The finite term $e(u_z)$ has a non trivial dependence on the velocity
$u_z$, and the numerical values are reported in the table.
For $u_z=0$ one recovers the static value already computed
in ref.\cite{fr,eh2}.

The renormalization of the velocity $\delta v_z$ is given, according
to eq.(\ref{eq:delv}), by:
\beqn
&&\frac{\delta v_z}{v_z}=\frac{g^2C_F}{16\pi^2}~\frac{1}{\pi}~[~
  v_0 \int \frac{d^3k}{\sqrt{(1+A)^2-1}}~
           \frac{2v_0z+v_z\sin k_z}{1-z+u_z\sin k_z}+
\nonumber\\
&&~~+\int \frac{d^3k}{\sqrt{(1+A)^2-1}}
\frac{[v_0^2~z(k)-v_z^2\cos^2(k_z/2)][z(k)-\cos k_z]}
     {[1-z+u_z\sin k_z]^2}~]
\label{eq:delv2}\eeqn
The tadpole graph does not contribute to the renormalization of the
velocity, because the heavy quark propagator does not enter inside the loop
and then it is not evaluated at large momenta.
The first term in eq.(\ref{eq:delv2}) is a lattice effect
while the second one has analog in the continuum
and the integrand vanishes for small $\vec{k}$
($z(k), \cos k_z\rightarrow 1$ for $\vec{k}\rightarrow 0$).

The velocity renormalization is a finite effect, because the infrared
divergences cancel between $X$ and $Y$,
and it can be written as:
\beqn
\frac{\delta v_z}{v_z}~=~\frac{g^2C_F}{16\pi^2}~c(u_z)
\label{eq:formu}\eeqn
The numerical values of $c(u_z)$ are reported in the table.
At $\beta=6$, formula (\ref{eq:formu}) gives a positive renormalization
of the velocity $\delta v_z$ that increases from $10$\% at $u=0.1$ up
to $18$\% for $u=0.7$.

Let us discuss now a method that allows in principle
a non-perturbative computation
of the renormalization of the velocity. We consider a specific example.

The correlator $G_M(t,u)$ of a meson $M$
composed of a light quark and an effective
quark with kinematical velocity $u$ behaves for $t\rightarrow\infty$
like:
\beqn
G_M(t,u)~\sim~\exp(-\epsilon(u) t)
\label{eq:meso}\eeqn
where $\epsilon (u)$ is the binding energy of a meson with velocity $u$
in the infinite mass limit. $\epsilon$ is not a physical quantity since
it contains the mass renormalization of the heavy quark $\delta M(u)$,
that is linearly divergent and has a complicate dependence on the velocity.

The correlator $G_H(t,u)$ of an hyperion composed
of light quarks and an effective
quark with velocity $u$  has a time
dependence analogous to that in eq.(\ref{eq:meso}),
with $\epsilon(u)$ replaced by
$\epsilon'(u)$, the hyperion binding energy.
By taking the ratio of the $2$-point functions, the mass renormalization
contribution $\delta M(u)$ to the binding energies cancels \cite{mms}:
\beqn
\frac{G_H(t,u)}{G_M(t,u)}~ \sim \exp[~-\Delta\epsilon(u)t~]
\eeqn
where $\Delta\epsilon(u)=\epsilon'(u)-\epsilon(u)$ is the difference of the
binding energies. $\Delta\epsilon(u)$ is a physical quantity
and, as such, satisfies the relativistic relation:
\beqn
\Delta\epsilon(u)=\gamma(u)\Delta\epsilon(0)
\label{eq:lore}\eeqn
where $\gamma(u)=1/\sqrt{1-u^2}$ is the Lorentz factor.

\noindent
By measuring  $\Delta\epsilon(u)$ and
$\Delta\epsilon(0)$ with numerical simulations,
one can derive by means of eq.(\ref{eq:lore}) the renormalized velocity $v_R$
of the heavy quark.

Let us consider now the vertex correction to the local heavy-heavy current
$J=\overline{h}_{v}\Gamma h_{v'}$.
The amplitude has already been reported in sec.4.
A general parametrization is the following:
\beqn
\delta V~=~\frac{g^2C_F}{16\pi^2}~
[~2(v\cdot v')r(v\cdot v')\ln(a\lambda)^2~+~d(v,v')~]
\label{eq:vert}\eeqn
where $r(x)=1/\sqrt{x^2-1}~\ln[x+\sqrt{x^2-1}]$.

\noindent
the logarithmic
term in eq.(\ref{eq:vert}), has already been computed in ref.\cite{llog};
it is a function of $v\cdot v'$, the only non
trivial invariant that can be constructed with the velocities $v$ and
$v'$ of the heavy quarks.
The finite term $d$ is not universal and in lattice regularization
depends separately on the components of $v$ and $v'$.
The constant $d$ has been evaluated numerically for the case of
one static quark, $u'=0$, and one quark moving along one axis
$\vec{u}=u_z\vec{z}$. The numerical values of $d(u)$
are reported in the table.

The one-loop matrix element of the current $J$
between heavy quark states is then given by:
\beqn
\langle h_{v}\mid J \mid h_{v'}\rangle &=&
 1+\frac{1}{2}\delta Z(v) +\frac{1}{2}\delta Z(v')+ \delta V(v,v')
\nonumber \\
&=&1+\frac{g^2C_F}{16\pi^2}
 [ 2(v\cdot v' ~r(v\cdot v') -1)\ln(a\lambda)^2 + f(v,v') ]~~~~
\eeqn
where we have used eqs.(\ref{eq:delz3},$~$\ref{eq:vert}) and
we have defined:
\beqn
  f(v,v')  =    \frac{1}{2}e(v) +\frac{1}{2}e(v') + d(v,v')
\eeqn
In the normalization point $v\cdot v'=1$ the anomalous dimension
of $J$ vanishes due to the conservation of the effective current
related to the flavor symmetry \cite{geo}.

For the case of an initial static quark $\vec{u}{~'}=0$ and a final quark
moving along the $z$ axis $\vec{u}=u\vec{z}$, the above
matrix element reduces to
\beqn
1+ \frac{g^2C_F}{16\pi^2}[~
         (\frac{1}{u}\ln\frac{1+u}{1-u}-2)\ln(a\lambda)^2 + f(u) ~]
\eeqn
The values of $f(u)$ are reported in the table at the end.

Let us discuss now the on-shell renormalization of the lattice effective theory
in the real space \cite{fr}, instead of in momentum space \cite{eh2} as
we have done up to now.
The two schemes differ on the lattice and the relation between them
has been clarified in \cite{mms}.

\noindent
Near the mass-shell (i.e. at large times) the self-energy
$\sum(k)$ can be written as:
\beqn
\sum(k)=-\delta M+\delta Z(~v_0(1-e^{-ik_4})+v_z\sin k_z~)+O(k^2)
\eeqn
where for simplicity we have neglected the velocity renormalization
$\delta v_z$ that is not important in this context.

\noindent
The bare propagator of the heavy quark on the lattice at order
$\alpha_s$, as function of time $t$ and momentum $\vec{k}$ is then given,
at large times, by:
\beqn
iH(t,\vec{k})&=&\int\frac{d k_4}{2\pi}e^{ik_4t}
\{~iH(k_4,\vec{k})+
\nonumber\\
&+&iH(k_4,\vec{k})[~
-\delta M +\delta Z(v_0(1-e^{-ik_4})+v_z\sin k_z)~]iH(k_4,\vec{k})~\}
\nonumber\\
&=&Z\frac{\theta (t)}{v_0}e^{-(t+1)\ln [1+u_z\sin k_z]}
\{1+\frac{-\delta M(t+1)}{v_0~(1+u_z\sin k_z)} \}
\nonumber\\
&=& Z \frac{\theta (t)}{v_0}\exp\{ -(t+1)\ln [1+u_z\sin k_z+\delta M/v_0]  \}
\label{eq:pro}\eeqn
where in the last line an exponentiation that is appropriate for
large $t$ has been done.

\noindent
In the continuum limit $a\rightarrow 0$,
the propagator (\ref{eq:pro}) reduces to:
\beqn
iH(t,\vec{k})=Z\frac{\theta(t)}{v_0}
       \exp[~-(t+1)(\delta M/v_0+\vec{u}\cdot\vec{k})~]
\label{eq:tpu}\eeqn
The renormalization conditions in momentum space
(\ref{eq:delm}-\ref{eq:delv}) imply then that
the field renormalization constant $Z$ is multiplied,
for an evolution of time $t$,
by the exponential with $t+1$ instead of $t$.

The evaluation of the Isgur-Wise function on the lattice requires
the computation of a $3$-point function $G$ containing two heavy quark
propagators.
According to eq.(\ref{eq:tpu}), the correlator $G$
contains the factors $\exp-(t+1)$ and $\exp-(t'+1)$ for the times
$t$ and $t'$ of the evolution of the two heavy quarks.

The bare propagator of the heavy quark with renormalization conditions
in the real space is given, in the limit $a\rightarrow 0$, by:
\beqn
Z'\frac{\theta(t)}{v_0}~\exp[-(\delta M'/v_0+\vec{u}\cdot\vec{k})t]
\label{eq:ti}\eeqn
Equating the expressions (\ref{eq:tpu}) and (\ref{eq:ti})
one gets the relation between the renormalization
constants in the two schemes:
\beqn
Z'~=~Z-\frac{\delta M}{ v_0 },~~~~~~~~\delta M'=\delta M
\eeqn
There is a finite difference in the wave function renormalization constants
$Z$ and $Z'$ because the mass renormalization $\delta M$ is linearly divergent,
i.e. $\delta M a$ does not vanish as $a\rightarrow 0$.

The renormalization constant of the operator $J$ in the real space
renormalization scheme is given by:
\beqn
\langle h_{v}\mid J \mid h_{v'}\rangle &=&
 1+\frac{1}{2}\delta Z'(v) +\frac{1}{2}\delta Z'(v')+ \delta V(v,v')
\nonumber \\
&=&1+\frac{g^2C_F}{16\pi^2}
 [~ 2(v\cdot v' ~r(v\cdot v') -1)\ln(a\lambda)^2 + f'(v,v') ~]
\nonumber\\
&=&1+ \frac{g^2C_F}{16\pi^2}[~
         (\frac{1}{u}\ln\frac{1+u}{1-u}-2)\ln(a\lambda)^2 + f'(u)~ ]
\label{eq:matr}\eeqn
where in the last line the case $\vec{u}{~'}=0$ and $\vec{u}=u\vec{z}$
has been considered. The values of $f'(u)$ are reported in the table.
Note that in the static limit $u=0$, $f'=0$, implying that the heavy-heavy
current $J$ is exactly conserved on the lattice with real space
renormalization conditions.

\section{Matching}

In this section we consider the matching of the lattice effective theory
with the full theory in the continuum \cite{ctsac}.

For a comparison of the theoretical rate of the decays
(\ref{eq:iwg}) with the
experimental one, it is necessary to convert the values of the form
factors computed with the lattice effective theory, to the values
in the original, 'true', theory.
In general, it may be useful to regularize the full theory with
a scheme different from the lattice one (and to renormalize it).
This implies that the lattice low-energy effective
theory and the 'target theory'
differ both in the high energy excitations and in the regularization.
The matching process is then conventionally divided in two steps:

\noindent
i) Matching between the bare amplitudes of the effective theory
   computed with lattice regularization, and the bare (or
   renormalized) amplitudes computed with a continuum
   regularization.

\noindent
ii) Matching between the amplitudes computed in the full and
    in the effective theory with the same regularization
    (or renormalization) scheme.

It is easy to see that both steps i) and ii), i.e. the whole
matching process, can be done in perturbation theory.

\noindent
In step i) we assume that infrared divergences are regulated in the
same way. Then, the two regularizations differ only in the precise
way in which they cut-off the high-momentum modes.
The difference of the amplitudes in the two
regularizations is related to hard parton effects, i.e. to partons
with momenta of order $1/a\gg\Lambda_{QCD}$.
Due to asymptotic freedom, this difference can
be safely computed in perturbation theory.

In step ii) one has an effective theory that is an expansion of the
full theory for momenta much less than the heavy quark mass $M_Q$.
At zero external momenta, i.e. in the matching point,
the loop amplitudes in the two theories differ only for virtual momenta of the
order or greater than $M_Q$.
Since $M_Q\gg\Lambda_{QCD}$ also in this case the difference can be computed
with perturbation theory.

In this paper we consider step i); step ii) has been considered in
ref.\cite{llog,compl}.

\noindent
The matrix element in eq.(\ref{eq:matr}) is given in the $\overline{MS}$
scheme by:
\beqn
\langle h_{v}\mid J \mid h_{v'}\rangle~=~1+ \frac{g^2C_F}{16\pi^2}
         (2-\frac{1}{u}\ln\frac{1+u}{1-u})\ln(\mu/\lambda)^2
\eeqn
The ratio of the $\overline{MS}$ matrix element divided by the lattice
matrix element gives the factor $Z_m$ by which
one has to multiply the values
of the lattice simulation, to get the $\overline{MS}$ values:
\beqn
Z_m~=~1+\frac{g^2C_F}{16\pi^2}
     [~(2-\frac{1}{u}\ln\frac{1+u}{1-u})\ln(\mu a)^2~-~f'(u)]~
\eeqn
We have considered the real space renormalization scheme (\ref{eq:ti}),
that appears more natural for the lattice matrix element of
the Isgur-Wise current.

\noindent
One sees explicitly that the dependence on the gluon mass
$\lambda$ cancels,
implying that soft contributions cancel in the matching.

For a numerical evaluation of $Z_m$ one has to select a value for
$\alpha_S$; at $\mu = 2$ GeV the lattice value is smaller than the
continuum one by a factor $2.7$. It is necessary to use a unique value
of $\alpha_S$, otherwise the matching constant is no longer an infrared safe
quantity.
One has to make a guess for the
higher orders of $g^2$ in $Z_m$.
By using the lattice value for $\beta = 6$ and taking $\mu=1/a$,
the matching constant $Z_m$ is $1$ at $u=0$ and decreases with the
velocity up to $0.95$ at $u=0.7$. With the continuum value of
$\alpha_s$, $Z_m=0.86$ at $u=0.7$.

We note that in the general case $\vec{v}\neq 0$ and $\vec{v}^{~'}\neq 0$,
the matching constant $Z_m$ depends separately on $v$ and $v'$, and
not only on $v\cdot v'$. The matrix elements of $J$ computed on the lattice
with different $v$ and $v'$ and the same $v\cdot v'$, must be
multiplied with different matching constants $Z_m(v,v')$ to cancel the effects
of the breaking of the Lorentz symmetry.

\section{Conclusions}

We have presented a fairly complete analysis of the effective theory
for heavy quarks in euclidean space. The results are encouraging:
the theory is consistent even though the energy spectrum is unbounded
from below. It is stable and has a sensible continuum limit that
reproduces the lowest order in $1/M$ of the full theory.
It is possible to compute the amplitude of an arbitrary process
of the effective theory in perturbation theory with a set of rules
that include (in addition to the usual Feynman rules) a prescription
for a contour integration over the energy.

Naive lattice regularization of the continuum euclidean action
leads to a duplication of the heavy quark species. This phenomenon
has not any significant effect in the applications to the phenomenology,
and it is then not necessary to add a Wilson-like term to the heavy
quark action.

The one-loop renormalization exhibits a series of phenomena that are
consequences both of the non-covariance of the lattice regularization
and of the fact that the effective theory selects a preferred direction
in the space-time, namely the heavy quark 4-velocity.

These peculiar effects of the lattice effective theory do not pose any
conceptual problem for the matching with the original relativistic
theory. Effective theories describing heavy quarks
with different velocities have to be considered as
different theories, due to the velocity super-selection rule that
appears in the infinite mass limit \cite{geo}.
All these effects of the lattice regularization
can be removed with velocity-dependent counter-terms.

\vskip .5 truecm
\appendix
\section{Subtraction of the infrared singularities}
\vskip .5truecm

In this appendix we give the formulas that have been used for the
subtraction of the infrared singularities of the loop
integrals for $a\lambda\rightarrow 0$,
and for the numerical evaluation of the integrals.

The singularity is isolated by subtracting and adding back
to the original integrand it's expansion for small momenta.
The difference is infrared finite and can be computed numerically
in the limit $a\lambda\rightarrow 0$.
The term added back is simpler than the original integrand and
can be integrated analytically.

In the limit of small momenta $\mid a k_i\mid\ll 1$ the singular term in
the wave function renormalization constant, eq(\ref{eq:delz2}), reduces to:
\beqn
h(\vec{k})~=~
\frac{1}{\pi v_0^2}~\frac{1}{ \sqrt{\vec{k}^{~2}+(a\lambda)^2} }
{}~\frac{1}{  [~\sqrt{\vec{k}^{~2}+(a\lambda)^2}+uk_z~]^2  }
\eeqn
where the usual phase-space factor $g^2C_F/16\pi^2$ has been
omitted.

\noindent
The function $h(\vec{k})$ can be easily integrated on a 3-sphere of radius $R$:
\beqn
\int {\rm d}^3k ~ h(\vec{k})
{}~=~2\ln(R/a\lambda)^2+\ln 16 -\frac{2}{u}\ln\frac{1+u}{1-u}
\label{eq:line}\eeqn

\noindent
Note that the linearized integral in eq.(\ref{eq:line}) depends on the velocity
$u$, since the domain of integration is a 3-sphere, that is not
$SO(4)$ invariant.

For the numerical computation,
formula (\ref{eq:delz2}) is then replaced by:
\beqn
&&\frac{1}{\pi}\int\frac{d^3k}{\sqrt{(1+A)^2-1}}~
           \frac{2v_0^2~z(k)+v_z^2u_z\sin k_z}{1-z(k)+u_z\sin k_z}+
\nonumber\\
        &+&\frac{1}{\pi}\int d^3k~\{~\frac{1}{\sqrt{(1+A)^2-1}}~
\frac{[v_0^2~z(k)-v_z^2\cos^2(k_z/2)][z(k)-u_z^2\cos k_z]}
     {[1-z(k)+u_z\sin k_z]^2}+
\nonumber\\
&-&\frac{1}{v_0^2}~\frac{\theta(R-\mid\vec{k}\mid)}
                { \mid\vec{k}\mid(\mid\vec{k}\mid +uk_z)^2 }~\}
+2\ln R^2 +\ln 16 -\frac{2}{u}\ln\frac{1+u}{1-u} -2\ln(a\lambda)^2~~~~
\eeqn
where the phase-space factor has been removed.

In the infrared limit $a k_i\rightarrow 0$,
the integral of the vertex correction (\ref{eq:iwvert}) reduces
to:
\beqn
I&=&\frac{-1}{\pi v_0v_0'}\int
\frac{{\rm d}^3 k}{ \sqrt{\vec{k}^{~2}+(a\lambda)^2} }
{}~\frac{v\cdot v'}{ (\sqrt{\vec{k}^{~2}+(a\lambda)^2}+\vec{u}\cdot\vec{k})~
          (\sqrt{\vec{k}^{~2}+(a\lambda)^2}+\vec{u'}\cdot\vec{k})  }
\nonumber\\
&=&\frac{-1}{\pi }\int {\rm d}^3 k
\frac{1}{ \vec{k}^{~2}+(a\lambda)^2}~
\frac{1}{ \sqrt{\vec{k}^{~2}+(a\lambda)^2}+\vec{u}\cdot\vec{k} }
\eeqn
with the usual factor removed.
In the last line we have taken $\vec{u}{~'}=0$.

\noindent
The integral above is not easy to compute analytically, and one has
to make a further subtraction. $I$ has the same infrared
singularity of the simpler integral
\beqn
L ~=~ \frac{-1}{\pi }\int {\rm d}^3 k
\frac{1}{ \mid\vec{k}\mid^{~2}+(a\lambda)^2}
\frac{1}{ \mid\vec{k}\mid+\vec{u}\cdot\vec{k} }
{}~=~-\frac{1}{u}\ln\frac{1+u}{1-u}\ln(R/a\lambda)^2
\eeqn
The vertex correction is then computed numerically by means of
the following formula:
\beqn
&&\frac{-1}{\pi }
\int d^3k  ~\{~    \frac{1}{ \sqrt{(1+A)^2-1} }
\frac{z(k)}{  [1-z(k)][1-z(k)+u_z\sin k_z]  }+
\nonumber\\
&&~~~~~~~~~~
  ~~~~~~~~~~~~-\frac{  \theta(R-\mid\vec{k}\mid)  }
      {  \mid\vec{k}\mid^2(\mid\vec{k}\mid+u k_z)  }  ~\}+
\nonumber\\
&&~~~~~~~-\delta(u)-\frac{1}{u}\ln\frac{1+u}{1-u}\ln R^2
+\frac{1}{u}\ln\frac{1+u}{1-u}\ln(a\lambda)^2
\eeqn
where $\delta(u)$ is a constant evaluated numerically:
\beqn
\delta(u) &=& L-I=
\nonumber\\
&=&\frac{2}{u}\int_0^{\infty}\frac{dk~k}{1+k^2}
\{~\ln\frac{\sqrt{1+k^2}+uk}{\sqrt{1+k^2}-uk}-\ln\frac{1+u}{1-u}~\}
\eeqn

\begin{center}
\section*{Table }
\section*{Numerical values of the renormalization constants}
\begin{tabular}{ccrccccrcc}
& & & & & & & \\
\hline
& & & & & &  &\\
$u$ & $x(u)$ & $e(u)$ & $e'(u)$ & $d(u)$ & $c(u)$ &  $f(u)$  & $f'(u)$  \\
& & & & & & & \\
& & & & & & & \\
\hline
& & & & & & & \\
& & & & & & & \\
.0   &  19.95  & 24.48   & 4.53  & -4.53   &    -    &   19.95   &  0.00  \\
.1   &  19.89  & 24.67   & 4.88  & -4.58   &  11.93  &   20.00   &  0.13  \\
.2   &  19.71  & 25.29   & 5.97  & -4.74   &  12.30  &   20.14   &  0.51  \\
.3   &  19.37  & 26.40   & 7.93  & -5.05   &  13.00  &   20.39   &  1.18  \\
.4   &  18.78  & 28.19   & 10.98 & -5.56   &  14.07  &   20.77   &  2.19  \\
.5   &  17.75  & 30.98   & 15.61 & -6.46   &  15.71  &   21.27   &  3.62  \\
.6   &  15.81  & 35.51   & 22.86 & -8.22   &  18.09  &   21.77   &  5.48  \\
.7   &  11.17  & 44.35   & 36.37 & -14.4   &  20.91  &   19.88   &  6.02  \\
& & & & & & & \\
\hline
& & & & & & & \\
& & & & & & & \\
\end{tabular}
\end{center}

\vskip .5truecm
\centerline{\bf Acknowledgement}
\vskip .5truecm

I wish to thank M. Crisafulli, R. Iengo, G. Martinelli, M. Masetti
and M. Testa for many useful discussions.

\newpage
\centerline{ \bf FIGURE CAPTIONS}
\vskip .5 truecm
\noindent
Fig.1: correlator of a system composed of an effective 'quark' $Q$
(see text) and a light antiquark $\overline{q}$ in the free case.
\vskip .5 truecm
\noindent
Fig.2: the same correlator as in fig.1 with one gluon exchange.
\vskip .5 truecm
\noindent
Fig.3: Wick rotation for the self-energy graph of the effective quark
$Q$ of fig.4. The crosses indicate the positions of the poles of
the effective quark $Q$, of the gluon $g$ and of the antigluon
$\overline{g}$.

a) Case of positive energy of $Q$: $\epsilon >0$

b) Case of $\epsilon <0$ and $\mid\epsilon\mid < E_{\lambda}$, where
   $E_{\lambda}=\sqrt{\vec{k}^{~2}+\lambda^2}$ is the gluon energy.

c) Case of $\epsilon < 0$ and $\mid\epsilon\mid > E_{\lambda}$.
\vskip .5 truecm
\noindent
Fig.4: self-energy graph of order $\alpha_S$ of the heavy quark.
\vskip .5 truecm
\noindent
Fig.5: tadpole graph of order $\alpha_S$ of the heavy quark.
\vskip .5 truecm
\noindent
Fig.6: vertex correction of order $\alpha_S$ of the current $J$
describing the transition of a heavy quark with velovity $v'$
into a heavy quark with velocity $v$.

\end{document}